\documentclass[prl, aps, twocolumn, groupedaddress,showpacs]{revtex4}

\usepackage{slashed}
\usepackage{graphicx}
\usepackage{subfigure}
\usepackage[usenames, dvipsnames]{color}
\usepackage{graphics}
\usepackage{hyperref}
\usepackage{bm}
\usepackage{amsfonts}
\hypersetup{backref,
colorlinks=true,
linkcolor=blue,
linktoc=page,
citecolor=blue}

\begin{document}

\title{High-Speed DC Magnetricity in Spinor Polariton Condensates}

\author{H. Ter\c{c}as}
\email{htercas@gmail.com}
\author{D. D. Solnyshkov}
\author{G. Malpuech}
\affiliation{Institut Pascal, PHOTON-N2, Clermont Universit\'e, Blaise Pascal University, CNRS,24 Avenue des Landais, 63177 Aubi\`ere Cedex, France}

\begin{abstract}
We investigate the spin dynamics of half-solitons in polariton Bose-Einstein condensates. Half-solitons, which behave as magnetic monopoles, can be accelerated in the presence of the effective magnetic field of the microcavity. We study the generation of DC magnetic currents in a gas of half-solitons. At low densities, the current is suppressed due to the dipolar  oscillations. At moderate densities, a magnetic current is recovered as a consequence of the collisions between the carriers. We show a deviation from Ohm's law due to the competition between magnetic dipoles and monopoles.
\end{abstract}
\maketitle

Since the original idea of Dirac \cite{Dirac1931}, magnetic monopoles have been one of the most important physical questions in quantum mechanics. In fact, ``real" elementary magnetic charges have not been observed up to now, despite long efforts to detect them \cite{Pinfold2010}. Recently, magnetically frustrated materials, or spin ices \cite{Bramwellprl, Castelnovo}, offered the possibility of investigating magnetic charge transport. Besides the substantial experimental evidence to support the existence of spin-ice magnetic monopoles \cite{Jaubert, Fennel, Morris, Kadowaki}, the measurement of the charge and current of magnetic monopoles has become possible \cite{Bramwell}. In fact, signatures of magnetic monopoles are present in other systems, such as nanowires \cite{NanowireScience} and spinor Bose-Einstein condensates \cite{Hivet}. Physically, these monopoles are elementary excitations in the system, or quasiparticles, a concept that is widely used in solid state physics to describe the behavior of carriers in the band structure \cite{Landau}. Modern electronics, for example, is successfully described in their terms. Quasiparticles differ from ``real" particles in the sense that they cannot exist outside the underlying medium.\par

An interesting example of quasiparticles are the so called half-solitons (HS) in spinor Bose-Einstein condensates (BECs). Half-solitons are stable localized excitations of spinor condensates with spin-anisotropic interactions \cite{Hivet, Flayac, Flayac2}. Recently, some of us have experimentally demonstrated that they behave like effective magnetic charges, being accelerated along applied effective magnetic fields \cite{Hivet}. Electricity, which is the basis of the modern world, is a current of electric charges in applied electric fields. By analogy, the motion of magnetic charges in magnetic field has been generally referred to as ``magnetricity" \cite{Bramwell}. Therefore, the idea of using HS to envisage magnetricity appears both natural and important. In particular, exciton-polariton condensates in semiconductor microcavities have been pointed out as extremely promising platforms to investigate magnetricity\cite{Bramwellnw}.

In this work, we present a theoretical study of polaritonic magnetricity: the collective motion of magnetic monopoles in the presence of an effective magnetic field. We show that at very low densities, the conductivity is suppressed due to dipole oscillations. At higher densities, when collisions between HS are more likely, the magnetic conductivity is optimal. For very dense gases, the conductivity decreases as a consequence of the collision time shortening. We also predict a deviation from the magnetic Ohm's law $j\propto H$ for moderate magnitudes of the applied field $H$. To confirm our predictions based on a kinetic model, we perform numerical simulations of realistic experimental configurations, where the DC conductivity can be effectively measured. Finally, we estimate the mobility of magnetic charges to 10$^7$ cm$^2$/Vs, an order of magnitude larger than the record value of the electronic mobility in graphene \cite{novoselov_NOBEL, bolotin}.

\begin{figure}[t!]
\includegraphics[scale=0.5]{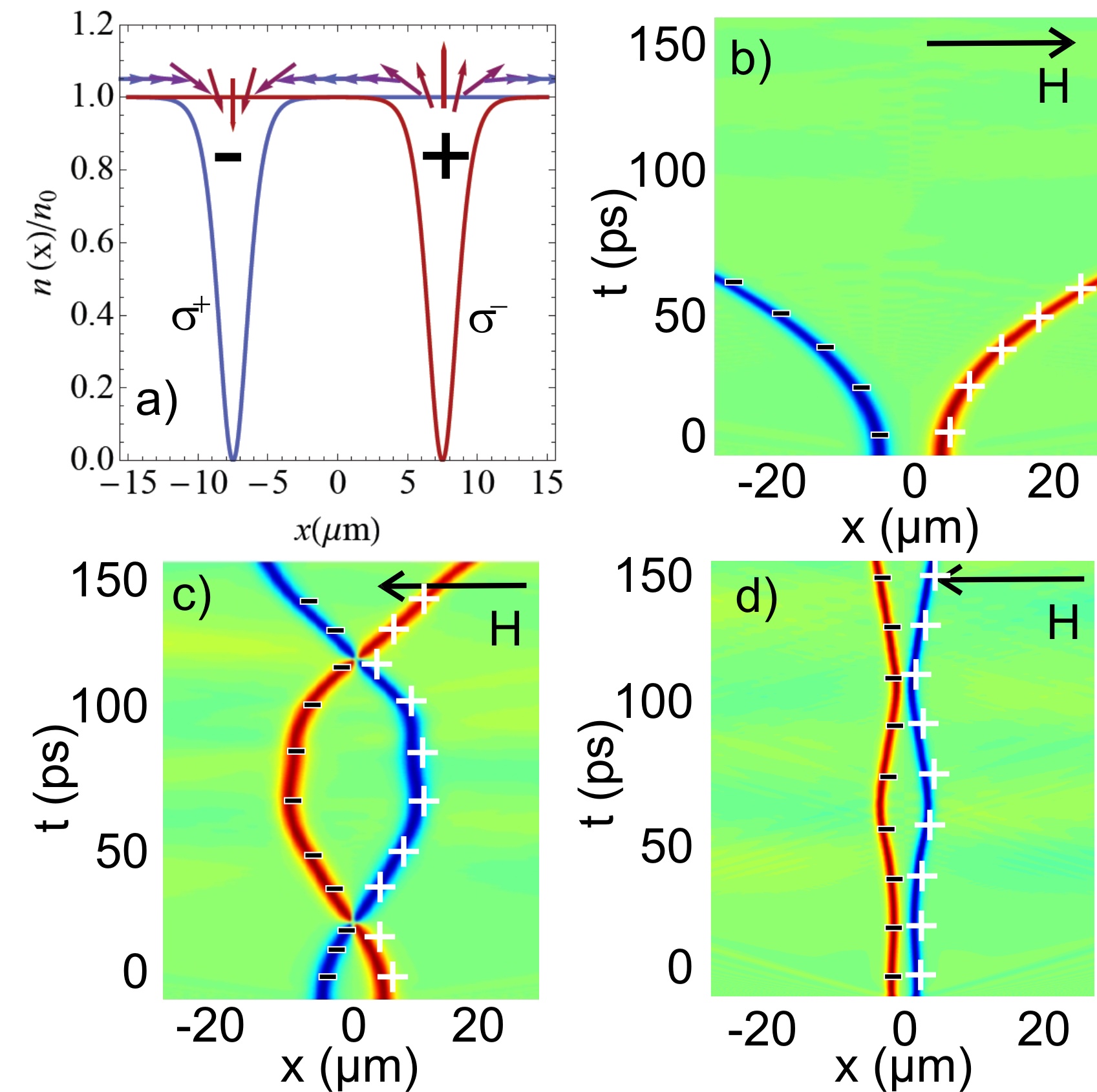}
\caption{(color online) Polarization degree $\rho_c$ for a pair of half-solitons. a) The red/dark gray (blue/light gray) line depicts the density profile of half-soliton in the $\sigma=-$ ($\sigma=+$) component. The arrows indicate the pseudospin $S_x$. Numerical time evolution of $\rho_c$ as extracted from Eq. (\ref{GP}) in the presence of a constant magnetic field $\mathbf{H}=\pm 10e_\mathbf{x}$ $\mu$eV (black arrows), showing the trajectories of two HSs: acceleration for $H>0$ (panel b)), and oscillations (panel c)) and bouncing (panel d)) for $H<0$. The symbols "+" and "-" indicate the sign of the magnetic charges.}
\label{fig1}
\end{figure}

\emph{Relativistic dynamics of half-solitons.} Exciton-polaritons are bosonic quasiparticles that result from strong light-matter coupling in semiconductor microcavities. Their most important properties in the framework of the present study are their capacity to form a condensate, their small effective mass and a very strong non-linearity. Exciton-polariton condensates in one-dimensional systems in the parabolic approximation can be described by the spinor Gross-Pitaevskii equation \cite{Shelykh, Carusotto}
\begin{equation}
i\hbar \frac{{\partial \psi _ \pm  }}
{{\partial t}} =  - \frac{{\hbar ^2 }}
{{2m}}\Delta \psi _ \pm   + \alpha _1 \left| {\psi _ \pm  } \right|^2 \psi _ \pm   + \alpha _2 \left| {\psi _ \mp  } \right|^2 \psi _ \pm  - H \psi_\mp,
\label{GP}
\end{equation}
Here,  $\mathbf{H}=H \mathbf{e}_x$ is the effective magnetic field along the $x$ direction due to the crystallographic anisotropy of the cavity \cite{Klopo}. Exciton-polaritons are also characterized by a strong spin-anisotropy (typically, $ \alpha_2 \approx  -0.2$-$0.1 \alpha_1$), which allows the existence of half-integer topological defects, such as half-solitons and half-vortices \cite{Lagoudakis}. Thus, the soliton solution of the spinor equation can be written in terms of a scalar soliton in a single spin component $\psi(x)=\sqrt{n_0}\tanh(x/\sqrt{2}\xi)$, with $\xi=\hbar/\sqrt{2\alpha_1m n_0}$ denoting the healing length \cite{pitaevskii, Santos, Frantzeskakis}
\begin{equation}
\psi_{+}=\sqrt{\frac{n_0}{2}}\left[i\frac{v}{c}+\frac{1}{\gamma}\tanh\left(\frac{x-y}{\sqrt{2}\xi\gamma}\right)\right], \quad
\psi_{-}=\sqrt{\frac{n_0}{2}}.
\label{HS1}
\end{equation}
Here, the half-soliton propagates with a velocity $v$, $y=vt+x_0$ is the soliton centroid, $c=\sqrt{\alpha_1n_0/m}$ is the sound speed and $\gamma=(1-v^2/c^2)^{-1/2}$ is the relativistic factor. This solution is characterized by a divergent in-plane pseudospin pattern $S_x=\textrm{Re}(\psi_+\psi_-^*)/2\simeq(n_0/2\gamma)\mathrm{sign}(y-x)$. Fig. \ref{fig1}a) shows the density and the pseudospin for two HS in opposite spin components. The magnetic charge is defined by analogy with Maxwell's equation $\rho=\bm \nabla\cdot \mathbf{S}$, and the charge of a single HS is $q=\pm n/2=\pm n_0/2\gamma$ (as shown by the symbols ``+" and ``-" in Fig. \ref{fig1}). Since the charge is defined by the in-plane pseudospin texture, it does not depend on the $\sigma_{\pm}$ component in which the HS appears.

The monopole dynamics of Eq. (\ref{HS1}) can be obtained by calculating the magnetic force $F_m=-n_0 H/2\gamma$ and the acceleration $a=n_0H/2M_0\gamma^2$, where $M_0=2\sqrt{2}n_0\xi m$ is the absolute value of the HS rest mass \cite{mass}. Integrating once, the velocity is $
v(t)=c\tanh\left(t/\tau_0\right)$ \cite{SolnyshkovMonop}, where $ \quad \tau_0=2M_0c/n_0 H$, which means that the soliton cannot be accelerated above the sound speed $c$. This trajectory is reproduced by numerical simulations of Eqs. (\ref{GP}).
\emph{Spin dynamics.} Let us now consider \emph{two} HS in different spin components, located at $\pm y/2$
\begin{equation}
\psi_{\pm}=\sqrt{\frac{n_0}{2}}\left[\pm i\frac{\dot y}{c}+\frac{1}{\gamma}\tanh\left(\frac{x\mp y/2}{\sqrt{2}\xi\gamma}\right)\right].
\label{HS2}
\end{equation}
The pseudospin texture is invariant with respect to the exchange of the two HS, $y\rightarrow -y$: for this particular solution, the spin field is divergent for the soliton on the right. Moreover, to assure the continuity of the phase, it is impossible to have two solitons of the same type (kink-kink) next to each other. Fig. \ref{fig1}, therefore, is the most general spin texture. When two solitons cross each other, the ``sign" of each monopole is inverted, i.e. the one located in the $\sigma_{-}$ projection, initially with a convergent texture, becomes divergent after crossing and vice-versa. In Fig. \ref{fig1} b), c), and d), we depict the temporal evolution of the solution (\ref{HS2}) by numerically computing the polarization degree $\rho_c=(n_+-n_-)/(n_+-n_-)$ in Eq. (\ref{GP}). Panel b) illustrates the simplest behavior: acceleration without crossing for $H>0$; panel c) illustrates the inversion of the charge (the `red', $\sigma_-$-soliton is initially accelerated to the left and then to the right). In this case, the two solitons undergo dipolar oscillations, forming a ``molecule", due to the changing in sign of the spin texture (charge). Panel d) depicts the bouncing of the two HS without the charge inversion, due to the interactions between spin components. We note the repulsive interaction between solitons for $\alpha_2<0$, as a consequence of their negative mass \cite{tercaswig}.

\begin{figure}[t!]
\includegraphics[scale=0.45]{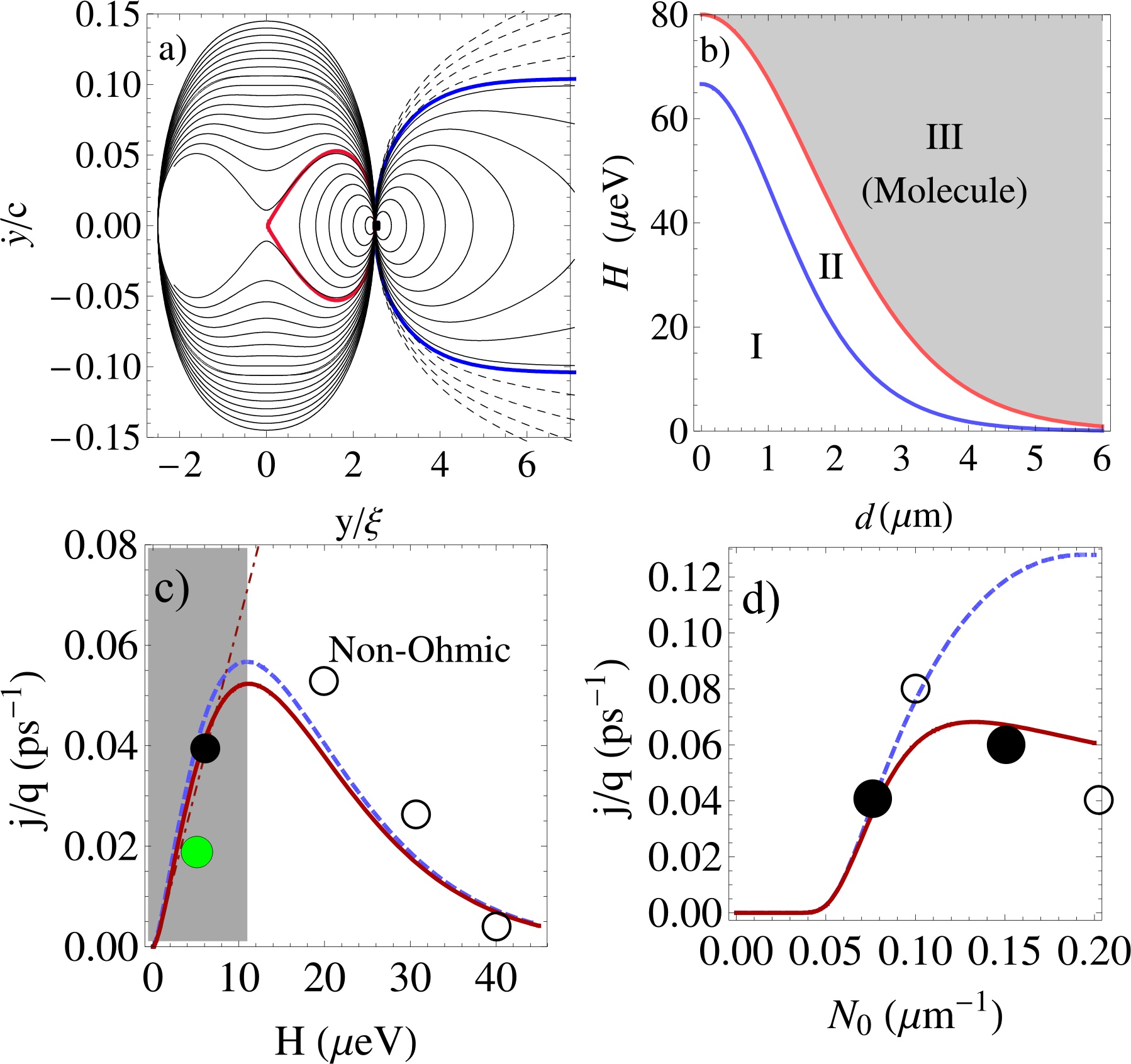}
\caption{(color online) a) Phase-space map for a pair of half-solitons initially separated by $d=2.5\xi$. Full (dashed) lines are obtained for $H>0$ ($H<0$). The thick line is obtained for $H=0$. The red/dark gray line is the separatrix between modes II and III. b) Magnitude of the critical fields $H_1$ (red/dark gray line) and $H_2$ (blue/light gray line) as a function of the initial separation $d$. c) DC magnetic current for $N_0=0.08\mu$m$^{-1}$. The shadow limits the Ohmic region.  d) Magnetic current as a function of density for $H=6\mu$eV. In panels c) and d), blue/light gray and red/dark gray solid lines are respectively obtained for $\alpha_2=0$ and $\alpha_2=-0.2\alpha_1$ and the dots represent the numerical results. Full dots are extracted from the simulations of Fig. \ref{fig3} b) and d) and the open dots are additional calculations (not shown). The full green/light gray dot is obtained for the configuration of Fig. \ref{fig3} d). Other parameters:  $m=5\times 10^{-5}m_e$, $\xi=1\mu$m and $n_0\sim 500\mu$m$^{-1}$.}
\label{fig2}
\end{figure}

We proceed to a variational analysis of the spin dynamics by using Eq. (\ref{HS2}) as an ansatz. The variational energy $E\left[y,\dot y\right]=\int \mathcal{E}~dx$, with $\mathcal{E}=\sum_{\sigma=\pm}\left[\frac{\hbar^2}{2m}\vert \psi_\sigma\vert^2 + \frac{1}{2}\alpha_1\left(\vert \psi_\sigma \vert^2-\frac{n_0}{2}\right)^2 \right]+\alpha_2\left(\vert \psi_+\vert^2\vert\psi_-\vert^2 -\frac{n}{2}\right)-H S_x$ representing the energy density \cite{pitaevskii, Santos}, is given by  
\begin{equation}
\begin{array}{c}
\displaystyle{E[y,\dot y]=\frac{4\sqrt{2}}{3}\left(1-\frac{\dot y^2}{c^2}\right)^{3/2}\alpha_1n_0^2\xi}+\sqrt{2}Hn_0\xi\zeta\coth\left(\zeta\right)+\\
\displaystyle{+\sqrt{2}\left(1-\frac{\dot y^2}{c^2}\right)^{3/2}\alpha_2n_0^2\xi \frac{\sinh\left(\zeta\right)-\zeta\cosh\left(\zeta\right)}{\sinh^3\left(\zeta\right)}},
\label{energy}
\end{array}
\end{equation}
where $\zeta=y(1-\dot y^2/c^2)^{1/2}/\sqrt{2}\xi$. The dynamics of a HS pair can then be calculated via the Hamilton equations, $\partial E/\partial y+d/dt(\partial E/\partial \dot y)=0$ and corresponds to that of a relativistic anharmonic oscillator. We restrict our discussion to the case of attractive inter-spin interaction, $\alpha_2=-0.2\alpha_1$. The results are summarized in Fig. \ref{fig2} a). If the solitons are accelerated away from each other ($H>0$), their trajectories correspond to open orbits in the phase-space; on the contrary, if accelerated towards each other, nonlinear oscillations of the HS molecule take place. Due to the competition between the short-range repulsion and the magnetic force, the system exhibits three types of oscillations, depending on the initial separation $d\equiv y(t=0)$: i) below the critical field $H_1$, defined through the condition $\partial E/\partial y\vert _{\dot y=0}=0$, repulsion dominates and the solitons bounce at distances larger than $d$ (mode I, also shown in Fig. \ref{fig1} d)); for  $H_1<H<H_2$, with $H_2$ defined by the contour $E\vert _{\dot y=0}=0$, dipolar oscillations possess an amplitude smaller than the initial separation $d$ (mode II); finally, for $H>H_2$, the solitons oscillate by crossing each other (mode III, also in Fig. \ref{fig1} c)). The critical fields $H_1$ and $H_2$ for the different oscillatory modes are illustrated in Fig. $\ref{fig2}$ b). For small amplitude oscillations, $d\lesssim \xi$, the dynamics is given by the equation $\ddot y+\omega^2 y=0$, and the oscillation frequency is given by
\begin{equation}
\omega=\frac{c_s}{\xi}\left(\frac{5H-4\alpha_2n_0}{15(H+4\alpha_1n_0+2\alpha_2n_0)}\right)^{1/2}.
\end{equation}
\emph{A set of half-solitons: the soliton gas.} To describe the dynamics of a gas of HSs, we postulate that the phase-space distributions $f^\pm(x,v,t)$ are governed by the following relativistic kinetic equations of the Vlasov type \cite{tercaswig}
\begin{equation}
\frac{\partial f^\pm}{\partial t}+v\frac{\partial f^\pm}{\partial x}+\frac{q(v)}{M(v)} H  \frac{\partial f^\pm}{\partial v}=I[f^\pm],
\label{solkin1}
\end{equation} 
where $q(v)/M(v)=n_0/2M_0\gamma^2$ is the relativistic charge/mass ratio. In order to estimate the transport properties of the system, in analogy with the Drude model for electrons in the presence of an electric field \cite{drude, sommerfeld}, we assume small departures from equilibrium, allowing for the collision integral $I[f^\pm]$ to be written in the relaxation-time approximation \cite{gantmakher}, $I[f^\pm]\simeq -(f^\pm-f_0^\pm)/\tau^\pm$, where $\tau^\pm$ is the relaxation time and $f_0^\pm$ is the phase-space equilibrium distribution. We define the total magnetic current as $j=j^+-j^-$, where $j^\sigma=\langle q(v) N_c v\rangle=\int q(v) N_c v f^\sigma dv $ and $N_c$ is the concentration of magnetic charges. For symmetry, the total current is given as $j=2j^+=-2j^-$, so we calculate the current associated with the $\sigma_+$ component, thus dropping the superscript in the equations above. From Eq. (\ref{solkin1}), the DC the magnetic current can be written as
\begin{equation}
j=\frac{N_cn_0^2\tau}{2 M_0}H\int v\left(1-\frac{v^2}{c^2}\right)\frac{\partial f_0}{\partial v}~dv.
\label{conductivity1}
\end{equation}
Eq. \ref{conductivity1} incorporates the relativistic behavior of HS, which implies a vanishing current near the sound speed $v\simeq c$. To estimate the collision time $\tau$, we make use of the Matthiessen's rule \cite{beaulac}: $
1/\tau=1/\tau_H+1/\tau_{\sigma,\sigma}+1/\tau_{\sigma,-\sigma},$ where $\tau_H$ is the collision rate induced by the field $H$; $\tau_{\sigma,\sigma}$ (resp. $\tau_{\sigma,-\sigma}$) represents the collision rate due to the short-range (but not contact) topological interaction between solitons of the same (resp. opposed) spin projection \cite{Santos, tercaswig}. A detailed derivation of $\tau$ is provided in the Supplemental Material \cite{suppl}. \par
As we have discussed, two-body dynamics is in competition with the collective behavior of the system. Thus, the concentration of available monopoles is not necessarily the same as that of the gas. To estimate the concentration of carriers, we extend Onsager's theory for the conduction of weak electrolytes \cite{suppl, Onsager}. Using the fermionic statistics of solitons, $f_0=N_0/(2v_F)\Theta(v_F-v)$ \cite{tercaswig}, Eq. (\ref{conductivity1}) yields
\begin{equation}
j=\frac{N_0n_0^2}{2M_0}\tau H\eta\left(1-\frac{v_F^2}{c^2}\right),
\label{current}
\end{equation}
with $\eta$ standing for the fraction of dissociated monopoles. The Fermi velocity of the gas, $v_F=\pi\hbar N_0/M_0$, is small compared to the sound speed for the case of polariton condensates, but it is not necessarily the case for cold atomic condensates$-$indeed, the same calculations could be performed for the latter$-$, for which we may have $n_0\xi\sim 1$. The features of Eq. (\ref{current}) are summarized in Fig. \ref{fig2} c) and d). For small values of the field, $\eta$ does not vary with $H$ and the DC current satisfies the Ohm's law $j\propto H$ (see Fig. \ref{fig2} c)). For moderate values of $H$, the system enters a non-ohmic regime, characterized by a negative conductivity, $\partial j/\partial H<0$. This behavior is qualitatively different from the deviation from the Ohmic response observed in spin-ices, where the conductivity monotonically increases with the applied field \cite{Bramwell}. The reason for such a difference resides in the fact that the soliton-pair dissociation energy depends on the density of the HS gas; besides, our system is one-dimensional and the jamming of carriers is more important than in spin-ices. In panel d) of Fig. \ref{fig2}, we plot the conductivity against the HS gas density. For very low densities, the transport is dominated by two-particle dynamics and the DC current is strongly suppressed. For higher densities, the response of the system is dictated by collisions. As a consequence, the magnetic conductivity reaches its maximum for moderate densities ($N_0\approx 0.12$ $\mu$m$^{-1}$ for $H=6~\mu$eV). The overall conductivity is suppressed if the interactions between HS in different spin projections are accounted for (about $30\%$ for $\alpha_2=-0.2\alpha_1$).

\begin{figure}[t!]
\includegraphics[scale=0.5]{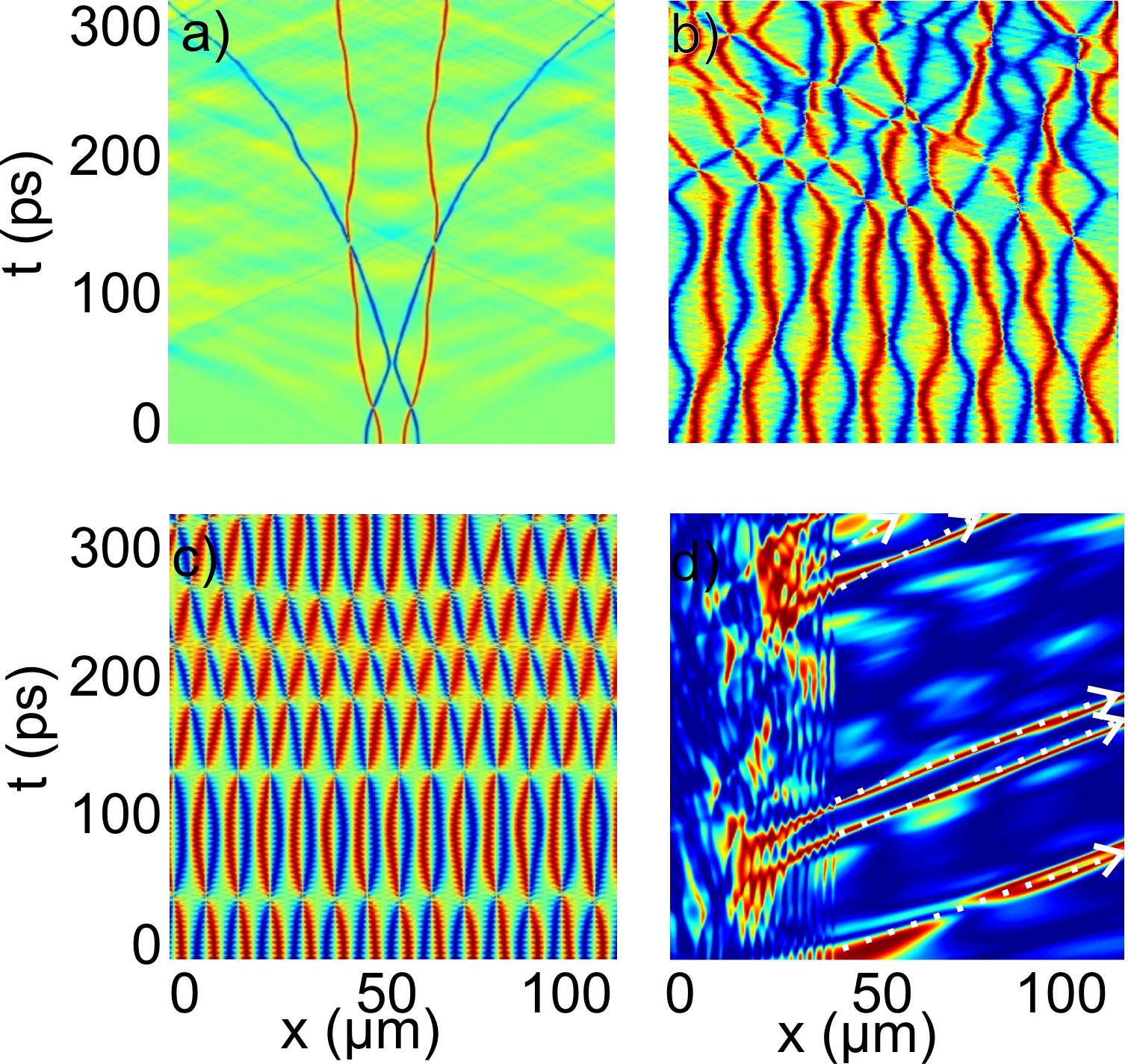}
\caption{(color online) a) Breakdown of oscillations due to the interactions between HS in the same component. b) A set of HS demonstrating the onset of magnetricity for $N_0\sim 0.08$ $\mu$m$^{-1}$. c) Suppression of conductivity due to short range interaction for $N_0\sim 0.15$ $\mu$m$^{-1}$. d) Extraction of half-solitons (red traces) from the trapping region ($\lambda=0.5~\mu$m,  $L=45\mu$m, $U_0=2$ meV and $\tau=30$) by applying a field of magnitude $H=5~\mu$eV. The white arrows are a guide for the eyes.}
\label{fig3}
\end{figure}

To corroborate the analytical predictions, we have performed numerical simulations to Eq. (\ref{GP}) with a gas of HS taken as initial condition. In Fig. \ref{fig3}, we illustrate the most relevant regimes of the magnetic current. In panel a), we observe the breaking of dipolar oscillations (or molecule dissociation) due to collisions between the solitons within the same component. For moderate values of density (panel b)), such collisions $-$ similarly to the Drude model for electron conduction $-$, lead to the appearance of a net current of magnetic charges (Ohmic response). Finally, for higher values of density, the conductivity is suppressed (panel c)), and the small-amplitude oscillations become the dominant mechanism. All these features are in qualitative agreement with the analytical estimates, as illustrated in Fig. \ref{fig2}. The size of the circles in panels c) and d) represent the precision of the numerical experiment. The deviation between the analytic theory and the numerical results stems in two important facts: i) the model neglects the acoustic radiation of solitons as they accelerate \cite{parker} (the acoustic Cherenkov effect); ii) the mean-field theory is expected to break down in one-dimensional systems. A suitable treatment of the transport properties in low dimensions must be done in the framework of Luttinger theory \cite{tercaswig, safi1, safi2}, but we leave this for a future work.         

To describe a more realistic experimental configuration, we simulate Eq. (\ref{GP}) for polaritons propagating in a one-dimensional cavity by adding i) two narrow Gaussian barriers described by the potentials $U_\pm\psi_\pm=\sum_{i=1}^2 U_0 \exp[-(x-x_i)^2/\lambda^2]\psi_\pm$, ii) a coherent, linearly polarized pump $P_{\pm} = P_0 e ^{i\left( {kx - \omega t} \right)}$ and iii) the finite life-time term $-i\hbar\psi_\pm/2\tau$. The two barriers, located at the positions $x_1=0$ and $x_2=L$, are strong enough to create and confine the HS gas. Initially, the magnetic field is absent, and the solitons remain trapped without escaping. Then, the effective magnetic field is switched on (it can be controlled externally \cite{GlazovAPL}), and we observe the extraction of HS from the confined region (red traces propagating to the right in Fig. \ref{fig3} d), showing linear polarization degree of the condensate. The corresponding current is marked in Fig. \ref{fig2} c) with a green circle. 

The monopole mobility can be directly estimated from our numerical simulations. To make a correct analogy with the usual definition of electronic mobility, we compare the potential energies corresponding to a fixed displacement. Indeed, a mobility of $10^6$ cm$^2$/Vs (a record value obtained in graphene \cite{novoselov_NOBEL, bolotin}) means that an electron is accelerated up to a speed a of $10^6$ cm/s in a field given by $1$ V in 1 cm. For the same distance, the displacement of a half-soliton corresponds to a magnetic energy of 10 eV (assuming $n_0\sim 2\times 10^2~\mu$m$^{-1}$ and a splitting of 5 $\mu$eV between the two spin states), while the velocity is $\sim 10^8~ $cm/s, which provides an equivalent mobility of $\mu=10^7$ cm$^2$/Vs. Such a high value is due to the extremely low polariton mass ($m<10^{-4}m_e$), which in turn is explained by an important photonic fraction.

To conclude, we have studied the magnetricity of a gas of half-solitons in spinor polariton condensates in the presence as effective DC magnetic field. We found that the monopole current deviates from the Ohmic response due to the competition of half-soliton oscillations and collisions. Record values of mobility can be expected due to the low polariton mass.

The authors acknowledge the support of the EU POLAPHEN, ANR Quandyde and GANEX projects.

\section{Supplemental}

In this supplemental Material, we include an explanation on the calculation of the DC magnetic current. In particular, we provide details on the computation of the soliton collision rate and on the determination of the monopole concentration - resulting from soliton-pair dissociation - by extending the Onsager's theory of weak  electrolytes.

\section{Collision time}

Within the relaxation-time approximation, the magnetic current described in the manuscript contains the collision time $\tau$, 

\begin{equation}
j=\frac{N_cn_0^2\tau}{2 M_0}H\int v\left(1-\frac{v^2}{c^2}\right)\frac{\partial f_0}{\partial v}~dv,
\label{conductivity1}
\end{equation}
for which different physical effects contribute. Neglecting the effect of impurities and other substract defects, we may identify three partial contributions: i) collisions induced by the magnetic field $\mathbf{H}=He_\mathbf{x}$, occurring at a rate $\tau_H$, ii) short-range collisions between solitons of the same spin ($\tau_{+,+}$) and iii) shot-range collisions between solitons of opposite spin projections ($\tau_{+,-}$). Making use of the Matthiesen's rule \cite{beaulac}, the collision time can thus be given by

\begin{equation}
\frac{1}{\tau}=\frac{1}{\tau_H}+\frac{1}{\tau_{+,+}}+\frac{1}{\tau_{+,-}}.
\label{coltime}
\end{equation}
The first term can be easily calculated by extending the original approach of Drude \cite{drude, sommerfeld}, from which the mean free  path $\ell$ between two solitons can be easily determined as
\begin{equation}
\frac{\ell}{2}=\int_0^{\tau_H} v(t) dt.
\label{tauH}
\end{equation} 
Using the expression for the relativistic velocity of solitons derived in the manuscript, $v(t)=c\tanh(t/\tau_0)$, with $\tau_0=2M_0c/n_0H$, and considering that the mean free  path and the density of the soliton gas are related as $\ell= 1/N_0$, we have
\begin{equation}
\tau_H=\tau_0\textrm{arccosh}\left[e^{-1/(2N_0\xi c\tau_0)}\right].
\label{matt}
\end{equation}

\begin{figure}[t!]
\includegraphics[scale=0.38]{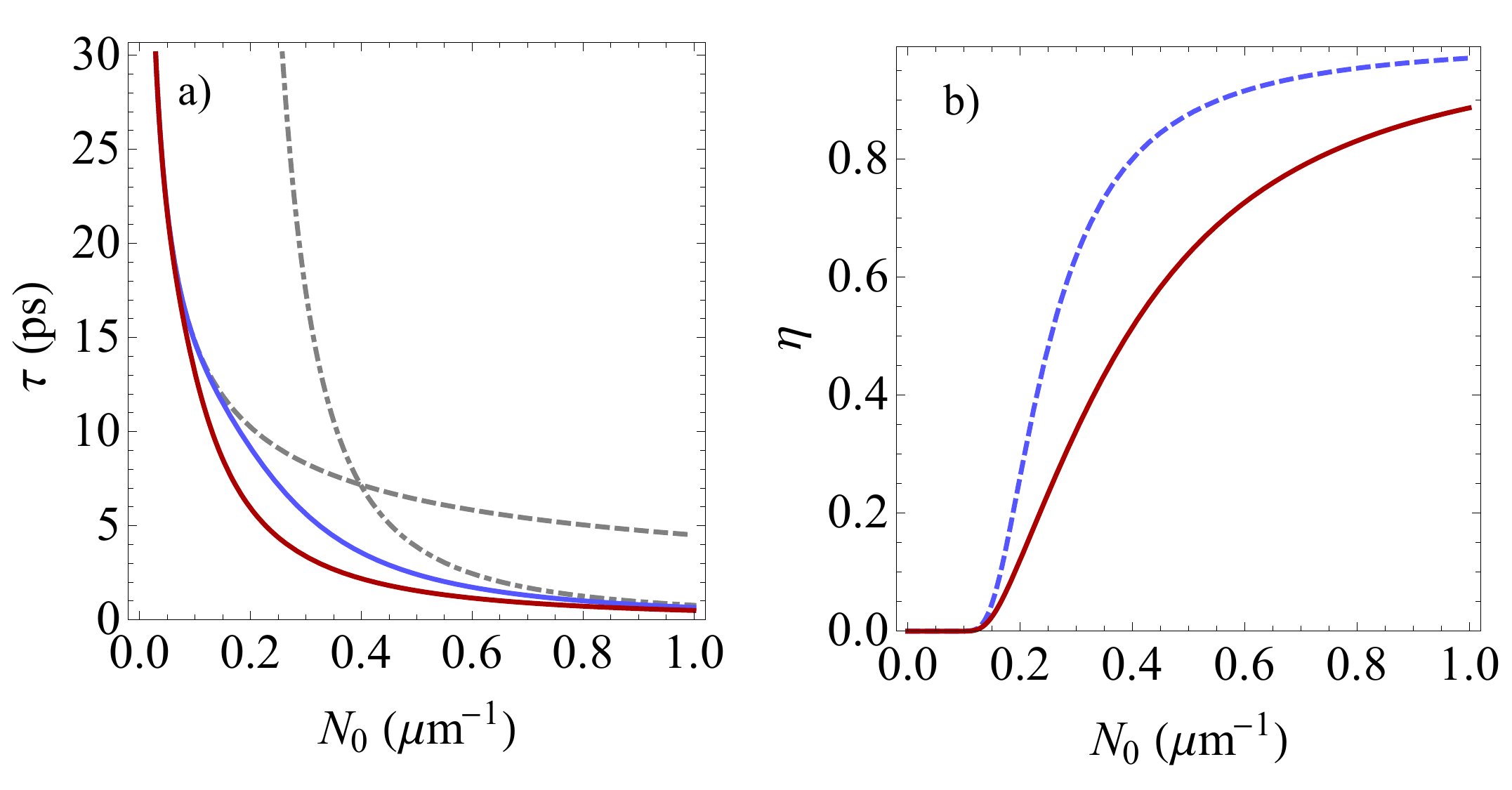}
\caption{(color online) a) Collision time of a balanced gas of half-solitons in the presence of a magnetic field $H=6\mu$eV. The dashed (dot-dashed) line depicts the rate $\tau_H$ ($\tau_{+,+}$). b) Relative monopole concentration $\eta$ as a function of the soliton gas density. In both panels, blue (light gray) and red (dark gray) solid lines are respectively obtained for $\alpha_2=0$ and $\alpha_2=-0.2\alpha_1$. We have used an experimentally accessible parameters for the condensate: $m=5\times 10^{-5}m_e$, $\xi=1~\mu$m and $n_0\sim 500~\mu$m$^{-1}$.}
\label{figS1}
\end{figure}

The contribution of intra-spin collisions to the collision rate can be estimated by analyzing a jamming situation. Let us assume two solitons in a line, initially at positions $x_1$ and $x_2$. Assuming completely elastic collisions, we can estimate that the relative kinetic energy is transformed into potential energy. In that case, we may write $E_{kin}^{i}=E_{pot}^{f}$, which reads
\begin{equation}
\frac{1}{2}M_0v^2=\frac{1}{2}M_0c^2\frac{1}{\sinh^2\left(d/\sqrt{2}\xi\right)},
\label{tau++}
\end{equation} 
where $d=\vert x_1-x_2\vert$. We have considered the short-range  potential resulting from the topological interaction between two half-solitons, previously considered in Ref. \cite{tercaswig}. Statistically, $\langle d\rangle=\ell$ and $\langle v \rangle=\ell/\tau_{+,+}$ and after averaging Eq. (\ref{tau++})
\begin{equation}
\tau_{+,+}=\frac{1}{N_0c}\sinh\left(\frac{1}{\sqrt{2}N_0\xi}\right).
\end{equation}
We proceed similarly to estimate the contribution due to inter-spin collisions. In our configuration, and in equilibrium, the distances between solitons of opposed spin projections are half of that between solitons of the same species, i.e. we consider an interstitial configuration. As a result, the mean distance between half-solitons of opposed spins is $\langle d \rangle=\ell/2$ and the collision rate reads
\begin{widetext}
\begin{equation}
\tau_{+,-}=\frac{\vert\Lambda\vert^{-1/2}}{2N_0 c}\left[\frac{\sinh^3\left(\frac{1}{2\sqrt{2}N_0 \xi}\right)}{\frac{1}{2\sqrt{2}N_0 \xi} \cosh\left(\frac{1}{2\sqrt{2}N_0 \xi}\right)-\sinh\left(\frac{1}{2\sqrt{2}N_0 \xi}\right)} \right]^{1/2},
\end{equation}
\end{widetext}
where $\Lambda=\alpha_2/\alpha_1$ is the relative inter-spin interaction strength. Here, we have replaced the rhs of Eq. (\ref{tau++}) by the topological potential of Ref. \cite{tercaswig}. For polaritons condensates, we have $\alpha_2\ll \alpha_1$, which means that the contribution to the reduced collision time $\tau$ is small and therefore intra-spin collisions dominate. In Fig. \ref{figS1} a) we plot the main features enclosed in Eq. (\ref{matt}). We observe that the low-density regime $N_0\xi\ll 1$ is determined by the acceleration in the presence of the magnetic field $H$. In the high density regime $N_0\xi\sim 1$, intra- and inter-spin collisions dominate.
\section{Dissociation rate and monopole concentration} 

As we have explained in the manuscript, the behavior of the half-soliton gas results from a competition between the spin dynamics and the collective kinetics of the particles. As a consequence, the number of carriers (monopoles) that effectively contribute to the DC magnetic current is not constant and strongly depends on both the magnetic field strength and on the gas density. In order to quantify such effect and to calculate the number of carriers $N_c$ in the system, we modify Onsager's theory, originally developed to model the electric response of weak electrolytes \cite{onsager}. The difference between the magnetic and inter-spin interaction potential (the later corresponding to a repulsion preventing the formation of a molecular state) of a pair of solitons separated by a distance $d$ is given by the two last terms of Eq. (4) of the manuscript by setting $\dot y=0$, thus reading
\begin{equation}
\begin{array}{lr}
\displaystyle{\Delta=Hn_0\xi\coth\left(\frac{d}{\sqrt{2}\xi}\right)}\\
\displaystyle{-\sqrt{2}\alpha_2n_0^2\xi\frac{\sinh\left(\frac{1}{\sqrt{2}\xi}\right)-d\cosh\left(\frac{d}{\sqrt{2}\xi}\right)/\sqrt{2}\xi}{\sinh^3\left(\frac{d}{\sqrt{2}\xi}\right)}}
\end{array}.
\end{equation}  
The probability of creating monopoles from the breaking of soliton molecules therefore increases with the ratio the molecular energy to the kinetic energy of the gas. At zero temperature, the concentration of monopoles contributing to the conduction can be (classically) given by the Boltzmann factor
\begin{equation}
\eta\equiv\frac{N_c}{N_0}=e^{\Delta/E_F},
\end{equation}
where $E_F=	\pi^2\hbar^2N_0/2M_0$ is the one-dimensional Fermi energy of the soliton gas, accounting for the statistical pressure of the gas. We notice that a detailed discussion about the fermionic nature of dark solitons has been given in Ref. \cite{tercaswig}. In fact, the phase singularity accompanying two-dark soliton wave function ``fermionizes" the condensate in one-dimensional systems. Finally, performing the substitution $d=1/2N_0$, we obtain the result discussed in the manuscript
\begin{equation}
j=\frac{N_0n_0^2}{2M_0}\tau H\eta\left(1-\frac{v_F^2}{c^2}\right).
\label{current}
\end{equation}
Notice that the Fermi velocity of the soliton gas, $v_F=\pi\hbar N_0/M_0$, is small compared to the sound speed for the case of polariton condensates, which means that the relativistic correction associated to the gas statistics may not be representative. However, it is not necessarily the case for cold atomic condensates, for which we may have $n_0\xi\sim 1$ and thus $v_F\sim c$ for experimentally accessible conditions.


\begin{thebibliography}{99}
\bibitem{mass}
The mass of each soliton is defined by the number of polaritons depleted from the condensate, $M=m\int \psi_\pm^*\psi_\pm dx=-2\sqrt{2}n_0\xi m/\gamma$, being therefore negative. $M_0$ is here defined as the magnitude of the rest mass.
\end{thebibliography}
\end{document}